\documentstyle[aps,epsf,multicol]{revtex}
\catcode`\@=11

\def\eqbegin         {  \begin{eqnarray}  }
\def\eqend           {  \end{eqnarray}  }
\def\beq{\begin{equation}}
\def\eeq{\end{equation}}

\def\del          { \partial }

\def\Z{{\bf Z}}

\def\hs_2{\hspace{2mm}}
\def\hs_3{\hspace{3mm}}
\def\hsf{\hspace{4mm}}

\title{Persistent Edge Current in the Fractional Quantum Hall Effect }
\author{Kazusumi Ino \thanks{e-mail:ino@kodama.issp.u-tokyo.ac.jp}}

\begin{document}
\maketitle
\begin{center}
{\it Institute for Solid State Physics, University of Tokyo,} \\
{\it Roppongi 7-22-1,  Minatoku,  Tokyo,  106, Japan} 
\end{center}
\thispagestyle{empty}
\begin{abstract}
We study the persistent edge current in  
 the fractional quantum Hall effect.
We give the grand partition functions for edge
excitations of hierarchical  states coupled 
 to an  Aharanov-Bohm flux and 
 derive the exact formula of the persistent edge current.   
For $m$-th hierarchical states with $m>1$,  it exhibits 
anomalous  oscillations in its flux dependence at low temperatures.  
The current as a function of flux goes to the sawtooth 
function with  period $\phi_0/m$ in the  
zero temperature limit.  
This phenomenon provides a new evidence for 
exotic condensation  in the fractional quantum Hall effect. 
We propose experiments of measuring the persistent edge current   
to  confirm the existence of the hierarchy.  
\end{abstract}


\begin{multicols}{2}
The predictions of the edge theory \cite{wen,kane} of the Fractional Quantum Hall 
(FQH) effect have been confirmed by tunneling experiments \cite{milli}
for $\nu=1/q$ ($q$ : odd integer). 
However a recent tunneling experiment \cite{gray} for
more general filling fractions  shows no evidence 
for the existence of the hierarchy \cite{jain,oldhier}. 
It is desirable to have other 
methods other than tunneling for the confirmation of the
hierarchy. Among others, 
the Aharanov-Bohm effect on the edge state may be a good
candidate.

Recently the Aharanov-Bohm effect (AB effect) on the 
edge state of the FQH  effect  attracted attentions
 \cite{chamon,geller}. 
 Especially, the amplitude of the persistent 
edge current  
induced by the AB flux was calculated for the $\nu=\frac{1}{q}$
Laughlin state by 
Geller et al \cite{geller}. It was shown that 
the current is periodic with period $\phi_0=hc/e$ as 
required by the general theorem of Byers-Young \cite{byer}. 
This result is  consistent with the microscopic 
study by Thouless and Gefen \cite{gefen}. As they argued, 
the existence of the fractionally charged quasiparticle 
is necessary for that period. 

In \cite{ino}, we investigated  the effect of 
pairing on the persistent edge current  
of quantum Hall analogs of BCS superconductor, so called 
paired FQH states \cite{halp,halrez,moore} in which 
the electrons with an even number of flux (composite 
fermions) are paired. 
We found anomalous oscillations of the current at 
low temperature. They 
 converge to the sawtooth function with 
period $\phi_0/2$ at zero temperature, 
which is an indication of  a pair condensation.  
This phenomenon was explained as follows: 
although the the bulk of paired states are BCS condensate 
of composite fermions, the condensation in the edge 
never occurs at finite temperature from the Mermin-Wagner 
theorem (no spontaneous symmetry breaking in 1+1 dimensions) 
\cite{mermin}.  
Thus the phenomenon provides an interesting bridge between 
the BCS condensation in 2+1 dimensions and the one 
in 1+1 dimension. 

On the other hand, it is known that  
composite bosonic objects of electron and an odd number of flux 
 (composite bosons) are condensated in the Laughlin states  and 
the hierarchical states \cite{girvin,zhanread}.  
It is naturally suggested 
that similar phenomenon  may occur in the  hierarchical states. 
In this paper, we study the persistent edge current 
in the  Jain' hierarchy \cite{jain}, which is by now a standard picture 
for the FQH effect. 
We address some other issues along with it.

Let us consider a single edge  of
a FQH state coupled to an AB flux.
For example, it is possible to change the strength of magnetic field
 for a FQH droplet within a given plateau.
In the composite boson picture, the situation is similar to  
flux piercing in a type II superconductor. 
Also we can consider the  boundary of an antidot in the Hall bar.

We first clarify the excitation structure on a single edge state 
 in the general hierarchical scheme \cite{read,wenzee}. 
Edge excitations of a state of 
$m$-th hierarchy involve $m$ species of 
chiral bosons $\varphi_i (i=1,\cdots,m).$ 
Their couplings are specified by 
an integer matrix $K$ \cite{read,wenzee}.
It is actually a metric of a lattice $\Gamma$, 
which is the lattice of the electron field 
\eqbegin 
e^{i{\bf v} \cdot {\bf \phi}}, \hspace{5mm} 
{\bf v}=\sum_{a=1}^{m}v^{a}{\bf e}^{a}, \hspace{3mm} v^{a} \in \Z. 
\label{lattice}
\eqend 
Here ${\bf e}_{a}$ $( a=1,\cdots,m )$ form a basis of the 
lattice $\Gamma $ with ${\bf e}_{a}{\bf e}_{b}=K_{ab}$. 
The filling fraction of the state is  given by 
\eqbegin
\nu = \sum_{a,b=1}^{m} (K^{-1})_{ab}. 
\eqend 
From the no monodromy condition between edge excitations and 
electrons,  
edge excitations are labeled by the dual lattice 
$\Gamma^{*} $ of $\Gamma$.  $\Gamma^{*}$ is spanned by 
the dual basis ${\bf e}^{*}_a$ with 
${\bf e}^{*}_a {\bf e}_{b} = \delta_{ab}$. 
Thus ${\bf w} \in \Gamma^{*}$ is a vector which 
satisfies ${\bf w}\cdot {\bf v} \in \Z. $
The energy  (conformal weight) $J$ and charge $Q$ of 
quasiparticle $e^{i{\bf w \varphi}}$  are given by
\eqbegin 
J = \frac{1}{2}{\bf w}\cdot {\bf w} 
, \hspace{3mm}Q={\bf t}\cdot{\bf w}.
\label{jq}
\eqend 
Here ${\bf t}$ is a real vector of charge unit with 
$t_a = 1/K_{aa}$.  
From these formulas, Fermi statistics of electrons 
requires $K_{aa}$ to be odd. 
By moding out the electrons, we see that fractionally charged quasiparticles are  
characterized  a lattice $\Gamma^{*}/\Gamma$. 
The lattice  $\Gamma^{*}/\Gamma$ has ${\rm det} K$ elements, 
which determines the topological order. 

The Jain's hierarchical scheme is obtained by 
taking  $K$ as 
\eqbegin 
K_{ab} = \delta_{ab} + s C_{ab}, 
\eqend 
where $s$ is a positive even integer and $C_{ab} =1 $ for 
$\forall a,b =1, \cdots,m$.  In this case, 
as ${\rm det} K = ms+1$, there are thus $ms+1$  sectors.  
From (\ref{jq}), the filling fraction is given by
\eqbegin 
\nu= \frac{m}{ms+1}. 
\eqend

The complete description of the edge 
excitations can be given  by  a  Virasoro character which 
corresponds to  the partition function of the grand ensemble of 
quasiparticles.  Summing up the characters for the points in $\Gamma^{*}$, 
we get the grand partition function as 
\eqbegin 
Z(\tau) = \frac{1}{\eta^{m}}
\sum_{\bf w\in \Gamma^{*}} {\rm exp}\left[ 2\pi i \tau 
\left(\frac{1}{2}{\bf w}\cdot{\bf w} \right) \right],
\label{hpf}
\eqend 
where $\eta$ is the Dedekind function 
$\eta(\tau)=x^{\frac{1}{24}}\prod_{n=1}^{\infty}(1-x^n),\hsf  
x=e^{2\pi i \tau}$ and  $\tau=i\frac{T_0}{T}$. 
$T_0=\frac{\hbar v}{ k_{B}L}$ is a temperature scale 
induced by $L$ the circumference of the edge state 
 and $v$ the  Fermi velocity of the edge modes.
For example, a Fermi velocity of $10^{6}$ cm/s and circumference of
$1\mu $ m yields $T_0 \sim  60$ mK.

Let us consider the 
effect of additional AB flux $\Phi$ for the hierarchical state. 
For that end, 
we note  that the AB flux only couples to  the charge degrees of freedom. 
Thus we need to  decompose $\Gamma^{*}$ into the neutral 
degrees of freedom and  the charge degrees of freedom explicitly. 

In $\Gamma^{*}$, the sublattice $\Gamma^{*}_{0}$ 
formed by points ${\bf t}\cdot{\bf w}=0$ is 
a $m-1$ dimensional lattice, which account for 
the neutral degrees of freedom.     
By a suitable $m-1$ dimensional modular 
transformation, the matrix $K_{ab}$ on $\Gamma^{*}_{0}$ 
 is equivalent to the Cartan matrix of SU($m$) \cite{read}. 
Actually, from (\ref{jq}), we see that 
there are $m(m-1)$ neutral excitations with $J =1$. 
As shown by Fr\"ohlich and Zee \cite{wenzee}, these excitations 
form the affine SU($m$) algebra of level $1$ and 
enlarge the chiral algebra. 
The states in $\Gamma^{*}$ 
are decomposed into $m$
 integrable representations of SU$(m)_1$, 
which can be labeled  by $a=1,\cdots,m$. They 
correspond to $a$-th antisymmetric tensor of SU($m$) 
respectively. Note that the electrons carry the quantum number
 of the  
fundamental representation of SU($m$), 
which corresponds to  Jain's idea of 
dividing the electron into $m$ species \cite{jain}.

To decompose $\Gamma^{*}$ into the  neutral SU$(m)_1$ degrees of freedom 
and the charge degrees of freedom,  
we consider  excitations which form a fundamental 
representation of SU($m$) with the charge   
$\frac{1}{m}\nu$.  
They are  'quark-like' excitations, generating  
all the excitations. The quantum number 
for a composite state of them is obtained by the fusion rules  
of SU$(m)_1$. The fusion rules of SU$(m)_1$ is given by 
\eqbegin 
a_1 \times a_2 = a_1+a_2  
\hspace{3mm} ({\it mod} \hspace{2mm}m),    
\label{fusionSUm}
\eqend 
which has a structure of $\Z/m\Z$. By using (\ref{fusionSUm}),   
 it is readily seen that 
a state which is in $a$ under SU$(m)_1$ 
has a charge $\frac{a}{m}\nu$ up to $\Z\nu$.

Under these quantum numbers of SU($m$) and U(1), 
we  can reorganize  the sum in (\ref{hpf}) as  \\
\eqbegin 
Z(\tau)=
\sum_{a=1}^{m}\chi_{a/m}(\nu\tau)
\chi^{{\rm SU}(m)_1}_{a}(\tau),  
\label{decom}
\eqend 
where $\chi^{{\rm SU}(m)_1}_{a}(\tau)$ is the 
character for representation $a$ of 
 affine SU$(m)_1$: \\  

$
\chi_a^{{\rm SU}(m)_1}(\tau)
= {\rm e}^{-\pi i \tau
\frac{a^2}{m}} \times    
$
\eqbegin 
 \times \frac{1}{\eta^{m-1}}\sum_{n_1+\cdots+n_m=a}
{\rm e}^{\pi i \tau (n_1^2+n_2^2+\cdots+n_m^2)}.
\eqend
Also $\chi_{a/m}(\tau)$ is defined as  
\eqbegin
\chi_{a/m} (\tau)=\frac{1}{\eta}
\sum_{n\in{\bf Z}}{\rm e}^{\pi \tau i
(n+a/m)^2}. 
\eqend
Similar decomposition of partition function has been given  
for the annulus case \cite{cappelli}.

Having separated the contribution from the charge degrees of 
freedom in $Z(\tau)$, let us consider the AB effect. 
An AB flux causes a spectral flow in charge spectrum
 which results in the change in $\chi_{a/m}(\nu\tau)$ as 
\eqbegin
\chi_{a/m}(\nu \tau,\phi)&=&
\frac{1}{\eta}\sum_{n\in{\bf Z}}{\rm e}^{\pi \tau i
\nu(n+a/m-\phi)^2}  \\ 
&\equiv&\frac{1}{\eta} \Theta\left[{ 0 \atop (m-a)/m}\right]
\left(\phi \vert\ t/\nu \right). 
\eqend 
where $\phi=\Phi/\phi_0$ with  $\phi_0=hc/e$  the unit 
flux quantum and $t=-\frac{1}{\tau}=i\frac{T}{T_0}$. 
$\Theta\left[{ a \atop b} \right]$ is the generalized $\Theta$ function 
\cite{mumford}. 
Thus the grand partition function with the AB flux becomes 
\eqbegin 
Z(\tau,\phi)=
\sum_{a=1}^{m} \frac{1}{\eta} \Theta\left[{ 0 \atop a/m}\right]
\left(\phi \vert\ t /\nu\right)
\chi^{{\rm SU}(m)_1}_{m-a}(\tau).  
\label{ABdecom}
\eqend
Obviously $Z(\tau,\phi)$ is a periodic function of $\phi$ with 
period $\phi_0$.

The exact formula for the persistent edge current is readily 
deduced from (\ref{ABdecom}).  
In general,  
the persistent current $I$ is defined by the following formula:  
\eqbegin 
I\equiv \frac{T}{\phi_0}
\frac{\del {\rm ln}Z(\tau,\phi)}{\del{\phi}}. 
\label{perfor}
\eqend 
From (\ref{perfor}) and (\ref{ABdecom}) 
we get the amplitude of persistent current induced by $\phi$ 
as 
\eqbegin
I(\tau,\phi)=\frac{T}{\phi_0}\frac{\sum_{\alpha=1}^{m}
\Theta'\left[{ 0 \atop a/m}\right]
\left(\phi \vert\ t/\nu \right)
\chi^{{\rm SU}(m)_1}_{m-a}(\tau)  
}{\sum_{\alpha=1}^{m}\Theta\left[{ 0\atop a/m}\right]
\left(\phi \vert\ t/\nu \right)
\chi^{{\rm SU}(m)_1}_{m-a}(\tau)}
\label{perJ}
\eqend  
where $\Theta'$ denotes the differentiation of $\Theta$ in $\phi$.
For example,  
the persistent current in the Laughlin state ($m=1$) is calculated to be   
\eqbegin 
I_{\rm Laugh} =\frac{2\pi T}{\phi_0}\sum_{n=0}^{\infty} (-1)^{n}  
\frac{\sin(2\pi n\phi)}{\sinh(nq\pi T/T_0)} 
\label{pers}
\eqend 
which is the formula obtained \cite{geller}.  
The periodicity of $I_{\rm Laugh}$ in $\Phi$ is $\phi_0$, which agrees with 
the general theorem of Byers and Yang \cite{byer}. 
This is due to the presence of the quasiparticle with 
a fractional charge as argued in Refs.\cite{geller}. 
Since there is no backscattering from impurities
 in the chiral Tomonaga-Luttinger liquid, the current has no reduction from 
impurities and therefore shows non-Fermi liquid dependence
 on the temperature.

\begin{figure}
\epsfxsize=7in
\epsfbox{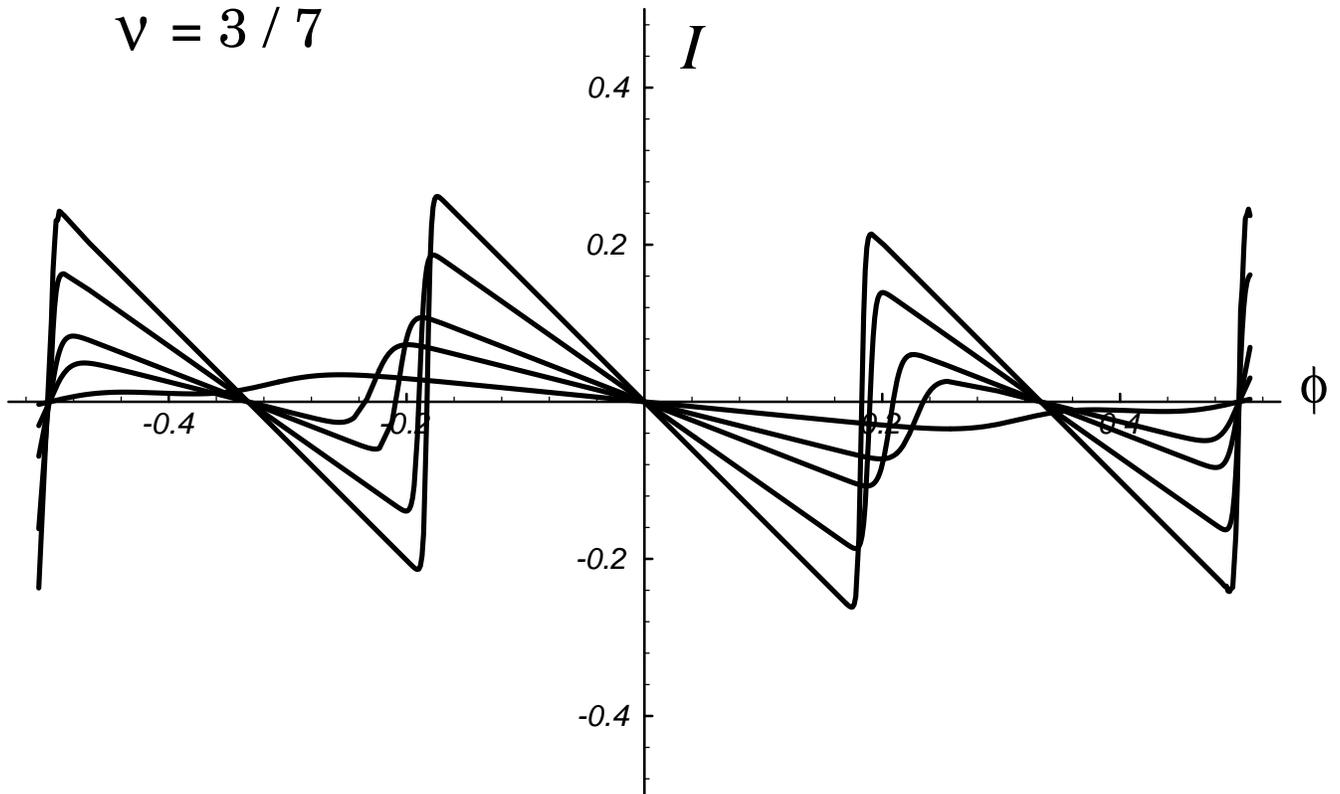}
\caption[flux3_7]
{ \    \    \  The flux dependence of the  persistent currents for the $\nu=3/7$ state $(m=3)$
at temperatures $T/T_0=0.29,0.23,0.21,0.18,0.17$. The currents
are measured in the unit 
$\frac{\nu ev}{2k_BL}$.}

\label{N3fig}
\end{figure}

\begin{figure}
\epsfxsize=7in
\epsfbox{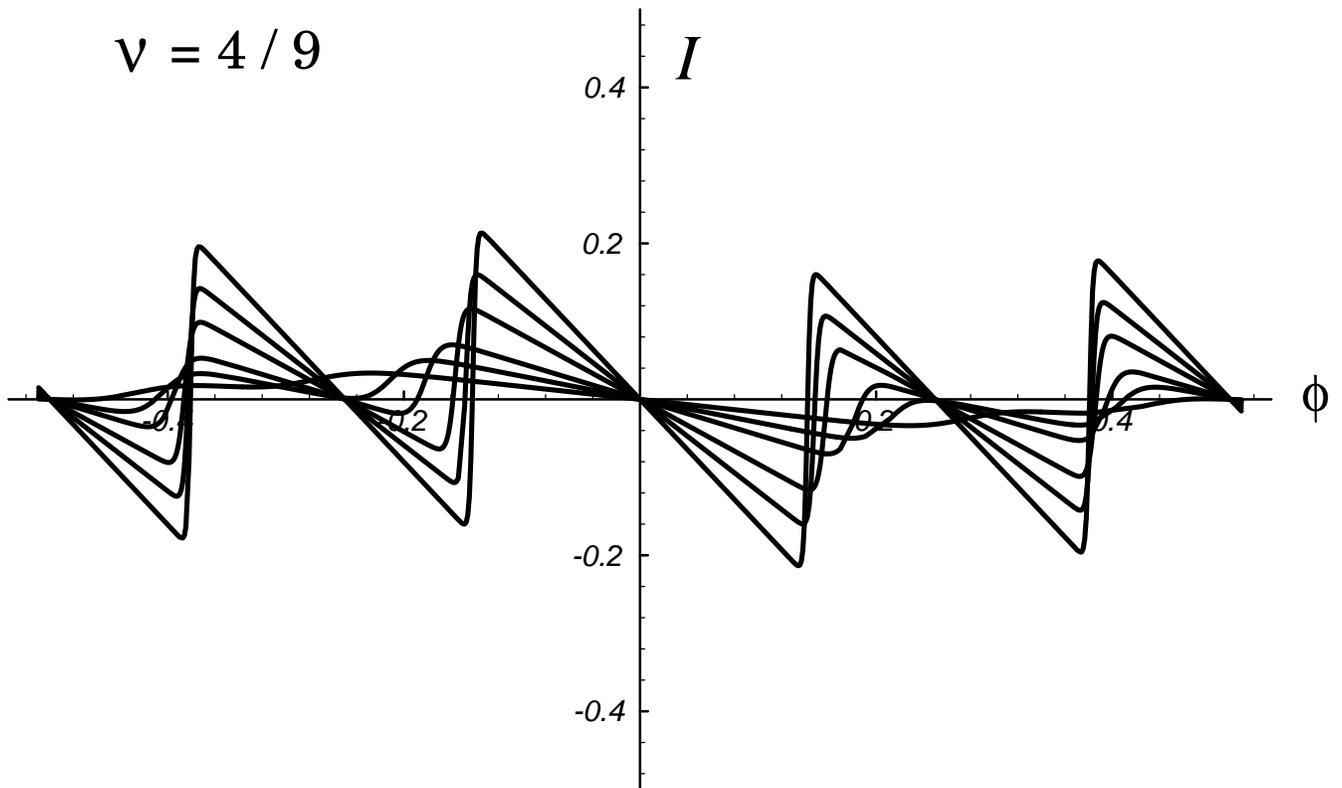}
\caption[flux4_9]
{ \   \    \  
The flux dependence of the  persistent currents for the $\nu=4/9$ state 
$(m=4)$ at temperatures $T/T_0=0.29,0.24,0.22,0.20,0.19,0.18$.  As the
temperature is lowered, the function approaches to a sawtooth function.}
\label{N4fig}
\end{figure}

For higher hierarchical states, the low temperature dependence turns out 
to be very different. 
The dependence of $I(\tau,\phi)$ is plotted in Fig.\ref{N3fig} for 
 $\nu=\frac{3}{7}$ $(m=3)$ 
case and Fig.\ref{N4fig} for $\nu=\frac{4}{9}$ 
$(m=4)$ case at various temperatures. 
We see that, as temperature is lowered, the shape of 
the graph of $I(\tau,\phi)$ continuously changes and 
exhibits anomalous oscillations. 
This is very similar to 
the case of paired FQH states \cite{ino}. However 
the shapes of the graphs  become close 
to the sawtooth function with  period 
$\phi_0/3$,  $\phi_0/4$  respectively. 
One observes similar behaviors 
for general $m$. In that case, the period of the 
sawtooth function at zero temperature is 
$\phi_0/m$ 
The convergence to the sawtooth function can be 
  analytically proved from the property of $\Theta$.  
We note that
$\Theta\left[{ 0\atop a/m}\right](\phi|t) = 
\theta_3(\phi+a/m|t)$ (Jacobi 3rd theta function) 
and $\theta_3$ is the fundamental solution of the heat equation in time $t$.
Thus, as the temperature is lowered, 
 $\Theta\left[{ 0 \atop  a/m}\right](\phi|\tau)
 $ is localized around 
$\phi=-a/m+\Z$    
 and the term proportional to 
$\Theta\left[{0\atop a/m}\right](\phi|t/\nu)$
  gives the dominant 
contribution to $I$ around $\phi=-a/m+\Z$.  
In the  zero temperature limit, it goes to 
\eqbegin 
I\rightarrow -\nu \frac{ev}{k_B L}(\phi-{a\over m}), 
\eqend 
for $-\frac{1}{2m}+{a\over m} < \phi < 
\frac{1}{2m}+{a\over m}, \ \ \
a \in {\bf Z}$. Note that only at zero temperature the period
$\phi_0/m$ is realized.  Although the period $\phi_0/2$ is 
common in the BCS superconductivity \cite{byer},  
the period $\phi_0/m$ for $m \ge 3$ is rather surprising. 
To our knowledge, such phenomenon has never been observed in experiments.

Physical explanation of this phenomenon may 
be given as in Ref.\cite{ino}: 
the condensation in the bulk drives the 
edge state to be condensated, but the 
Mermin-Wagner theorem tells that the 
condensation can occur only at zero temperature. 
In hierarchical states, the object which 
supposed to be condensated is the composite boson \cite{girvin}, 
 comprised from the electron and an odd number of flux quantum. 
The phenomenon we have found 
provides  a new evidence for the exotic condensation in FQH states.  
The phenomenon  clearly shows that the order parameter in hierarchical state 
has charge  $m$.  Thus the order parameter should be the
$m$-th antisymmetric tensor invariant of SU$(m)$ composed from 
$m$ composite bosons.
Although the SU$(m)$ invariant formulation 
of the order parameter and Landau-Ginzburg
theory with SU$(m)$ Chern-Simons gauge fields  are not known, 
the property of the persistent edge current suggests 
such formulation exists. 
On the other hand, as sketched  by Moore and Read \cite{moore}, 
 the extending fields in the {\it bulk} rational conformal field theory 
 of FQH state have  properties of  the order parameter in FQH state.  
Our results may serve as the evidence of 
their suggestion from the {\it edge} conformal field theory. 
Chern-Simons theory already has information of condensation.

The persistent edge current should be related to 
the stability of the FQH state since, in the composite 
boson picture,  the development of 
plateau arises from the flux piercing which is 
similar to the vortex state of a type II superconductor. 
 The flux piercing  accompanies 
the persistent edge current as its linear response.  
For a $m$-th hierarchical state, 
the flux dependence of the current $I$ at very low
temperature is close to the sawtooth function with the period of 
 $1/m$ to the one for  most stable Laughlin states. 
Accordingly, the maximal amplitude of the current is reduced.
It is expected that, at the fluxes where the direction of 
the current is reversed, the state may be unstabilized, which 
results in a $1/m$ narrower plateau.  
This is consistent with the general claims on hierarchy: 
as the rank of the hierarchy becomes higher, the state is more unstable 
\cite{oldhier,jain}.  
Thus the reduction of the amplitude and period of the persistent edge
current at zero temperature  qualitatively 
explains the narrower width of  plateau for higher hierarchical states.

Experimentally, this phenomenon provides unambiguous means to 
detect  the hierarchy of a given FQH state. 
 In this phenomenon  the hierarchy in the FQH state 
manifests itself in a physical quantity 
in an unambiguous way. 

Finally we would like to comment on states with $\nu=m/(ms-1)$. 
In that case, it is known that the edge state is not 
decomposed into chiral and antichiral sectors \cite{cappelli}. 
Formally the formula for $\nu=m/(ms-1)$ can be obtained by 
replacing $\chi^{{\rm SU}(m)_1}_{a}(\tau)$ by
$\overline{\chi^{{\rm SU}(m)_1}_{a}(\tau)}$ in (\ref{ABdecom}). 
However, as  
 the current has a backscattering and reduction from impurities, 
the formula is not valid at finite temperature.

In conclusion,  we investigated the persistent edge current in
the hierarchical FQH states.  
We found that the persistent edge current $I$ is a
periodic function of an AB flux  with period $\phi_0$ and that
at zero temperature its period changes to $\phi_0/m$. 
We also gave a new explanation of the stability of hierarchical
states from this property. 
The experimental observation of the persistent edge current 
enables one to  determine the level $m$ of the hierarchical
 FQH state  in an unambiguous way.

{\it Acknowledgement.} 
We would like to thank M. Kohmoto, D. Lidsky, S. Murakami, 
J. Shiraishi for various comments and suggestions. 
We also would like to thank K.Imura for  
a discussion on the $\nu=\frac{2}{5}$ case 
 at the initial stage of this work.

\vskip 0.6in

\def\NP{{ Nucl. Phys.\ }}
\def\PRL{{ Phys. Rev. Lett.\ }}
\def\PL{{ Phys. Lett.\ }}
\def\PR{{ Phys. Rev.\ }}
\def\CMP{{ Comm. Math. Phys.\ }}
\def\IJMP{{ Int. J. Mod. Phys.\ }}
\def\MPL{{ Mod. Phys. Lett.\ }}
\def\RMP{{ Rev. Mod. Phys.\ }}
\def\AP{{ Ann. Phys. (NY)\ }}

\end{multicols}

\newpage 

\end{document}